\documentclass[review,3p,times,sort&compress]{elsarticle}
\usepackage[colorlinks,citecolor=magenta,linkcolor=blue]{hyperref}
\usepackage{graphicx}
\usepackage{natbib}
\usepackage{amsmath,amssymb}
\usepackage{txfonts,bm,listings,xcolor}

\newcounter{bla}

\lstdefinelanguage{Julia}{morekeywords={using,
    push, open, do, for, in, eachindex,
    end, Dict, String, Any, im, printf,
    LOAD_PATH, ENV},
    sensitive=true,
    morecomment=[l]{\#},
    morestring=[b]",
}

\lstdefinelanguage{TOML}{morekeywords={Test,solver,
    ptype,ktype,grid,mesh,ngrid,nmesh,ntest,wmax,wmin,pmax,pmin,beta,noise,offdiag,lpeak,
    Solver,method,stype,nalph,alpha,ratio,blur},
    sensitive=true,
    morecomment=[l]{\#},
    morestring=[b]",
}

\journal{Computer Physics Communications}

\begin{document}

\begin{frontmatter}

\title{ACTest: A testing toolkit for analytic continuation methods and codes}

\author[a]{Li Huang\corref{author}}

\cortext[author] {Corresponding author.\\\textit{E-mail address:} huangli@caep.cn}
\address[a]{Science and Technology on Surface Physics and Chemistry Laboratory, P.O. Box 9-35, Jiangyou 621908, China}

\begin{abstract}
ACTest is an open-source toolkit developed in the Julia language. Its central goal is to automatically establish analytic continuation testing datasets, which include a large number of spectral functions and the corresponding Green's functions. These datasets can be used to benchmark various analytic continuation methods and codes. In ACTest, the spectral functions are constructed by a superposition of randomly generated Gaussian, Lorentzian, $\delta$-like, rectangular, and Rise-And-Decay peaks. The spectra can be positive definite or non-positive definite. The corresponding energy grids can be linear or non-linear. ACTest supports both fermionic and bosonic Green's functions on either imaginary time or Matsubara frequency axes. Artificial noise can be superimposed on the synthetic Green's functions to simulate realistic Green's functions obtained by quantum Monte Carlo calculations. ACTest includes a standard testing dataset, namely ACT100. This built-in dataset contains 100 testing cases that cover representative analytic continuation scenarios. Now ACTest is fully integrated with the ACFlow toolkit. It can directly invoke the analytic continuation methods as implemented in the ACFlow toolkit for calculations, analyze calculated results, and evaluate computational efficiency and accuracy. ACTest comprises many examples and detailed documentation. The purpose of this paper is to introduce the major features and usages of the ACTest toolkit. The benchmark results on the ACT100 dataset for the maximum entropy method, which is probably the most popular analytic continuation method, are also presented.
\end{abstract}


\end{frontmatter}

\noindent {\bf PROGRAM SUMMARY}\\

\begin{small}
\noindent
{\em Program Title:} ACTest \\
{\em CPC Library link to program files:} (to be added by Technical Editor) \\
{\em Developer's repository link:} https://github.com/huangli712/ACTest \\
{\em Code Ocean capsule:} (to be added by Technical Editor)\\
{\em Licensing provisions (please choose one):} GPLv3 \\
{\em Programming language:} Julia \\
{\em Supplementary material:} \\
  % Fill in if necessary, otherwise leave out.
{\em Journal reference of previous version:}* \\
  %Only required for a New Version summary, otherwise leave out.
{\em Does the new version supersede the previous version?:}* \\
  %Only required for a New Version summary, otherwise leave out.
{\em Reasons for the new version:}* \\
  %Only required for a New Version summary, otherwise leave out.
{\em Summary of revisions:}* \\
  %Only required for a New Version summary, otherwise leave out.
{\em Nature of problem (approx. 50-250 words):}
Analytic continuation is an essential step in quantum many-body computations. It enables the extraction of observable spectral functions from imaginary time or Matsubara frequency Green's functions. Though numerous methods for analytic continuation were proposed, they have not been systematically and fairly benchmarked due to the lack of testing software and datasets. \\
{\em Solution method (approx. 50-250 words):}
At first, a few peaks (features) are generated randomly. Then, they are superimposed to form the spectral function. Finally, the Green's function is reconstructed from the spectral function via the Laplace transformation. This procedure allows for the generation of lots of spectral functions and corresponding Green's functions, creating datasets for testing analytic continuation methods and codes. \\
\end{small}

\tableofcontents

\section{Introduction\label{sec:intro}}

In quantum many-body computations, analytic continuation usually plays a vital role. Its objective is to convert Green's function $G$ from imaginary time ($\tau$) or Matsubara frequency ($i\omega_n$) axis to real time ($t$) or real frequency ($\omega$) axis, extracting spectral function $A(\omega)$ for comparison with experimental data~\cite{many_body_book,many_body_book_2016,RevModPhys.68.13}. From a mathematical perspective, $G(\tau)$ or $G(i\omega_n)$ is related to $A(\omega)$ through the well-known Laplace transformation. Given $G(\tau)$ or $G(i\omega_n)$, analytic continuation calculation is essentially finding the associated $A(\omega)$ that satisfies the Laplace transformation. This is a typical inverse problem. Note that the solution, i.e., $A(\omega)$, is particularly sensitive to numerical fluctuations or noises in the input data, i.e., $G(\tau)$ or $G(i\omega_n)$. This poses a severe challenge to the analytic continuation methods~\cite{ASAKAWA2001459,PhysRevB.34.4744,PhysRevLett.55.1204}.

Over the past few decades, people have developed many analytic continuation methods, including the Pad\'{e} approximation~\cite{PhysRevB.87.245135,PhysRevB.93.075104,Vidberg1977,PhysRevD.96.036002}, maximum entropy method~\cite{JARRELL1996133,PhysRevB.41.2380,PhysRevB.44.6011,PhysRevB.81.155107}, stochastic analytic continuation~\cite{beach2004,PhysRevB.57.10287,PhysRevE.94.063308,PhysRevX.7.041072,SHAO20231,PhysRevB.102.035114,PhysRevB.101.085111,PhysRevB.78.174429,PhysRevE.81.056701,PhysRevB.76.035115}, stochastic optimization method~\cite{PhysRevB.62.6317,PhysRevB.95.014102}, stochastic pole expansion~\cite{PhysRevB.108.235143,PhysRevD.109.054508}, Nevanlinna analytical continuation~\cite{PhysRevLett.126.056402,PhysRevB.104.165111}, sparse modeling~\cite{PhysRevE.95.061302,PhysRevB.105.035139}, causal projections~\cite{PhysRevB.107.075151}, Prony fits~\cite{Ying2022,YING2022111549,PhysRevB.110.035154}, and machine learning-aided methods~\cite{PhysRevLett.124.056401,PhysRevB.98.245101,PhysRevB.105.075112,PhysRevResearch.4.043082,Yao_2022,Arsenault_2017,PhysRevE.106.025312}, to name a few. Unfortunately, nowadays there is no perfect and universal method for solving analytic continuation problems. The existing methods have their own advantages and disadvantages, as well as their scopes of applications. We need a quantitative comparison standard to determine their relative merits. In addition, people have developed several analytic continuation toolkits, such as ACFlow~\cite{Huang:2022}, $\Omega$Maxent~\cite{PhysRevE.94.023303}, ana\_cont~\cite{KAUFMANN2023108519}, Nevanlnna.jl~\cite{10.21468/SciPostPhysCodeb.19}, maxent (in ALPSCore)~\cite{LEVY2017149}, Nevanlinna (in TRIQS)~\cite{ISKAKOV2024109299}, SOM (in TRIQS)~\cite{KRIVENKO2019166,KRIVENKO2022108491}, Stoch (in ALF)~\cite{10.21468/SciPostPhysCodeb.1}, SmoQyDEAC.jl~\cite{10.21468/SciPostPhysCodeb.39}, etc. Are these analytic continuation toolkits reliable? Can their accuracy meet the requirements? These are open questions that need to be addressed.

To answer or solve the aforementioned questions, we would like to introduce an open-source toolkit ACTest in this paper. ACTest can randomly generate a large number of $A(\omega)$, along with the corresponding $G(\tau)$ or $G(i\omega_n)$. It is worth emphasizing that $G(\tau)$ or $G(i\omega_n)$ should be supplemented with artificial noises to mimic input data from realistic quantum many-body calculations. The synthetic $A(\omega)$ and $G(\tau)$ [or $G(i\omega_n)$] can be used to benchmark the existing or newly developed analytic continuation methods, examining their computational accuracies and efficiencies. The ACTest toolkit comes with a standard dataset containing 100 spectral functions, known as ACT100. This built-in dataset can be used to assess the merits of different analytic continuation methods in a relatively fair, reproducible, and quantitative manner. People can also employ the ACTest toolkit to create large datasets, which can serve as training and testing datasets for machine learning-aided methods~\cite{PhysRevLett.124.056401,PhysRevB.98.245101,PhysRevB.105.075112,PhysRevResearch.4.043082,Yao_2022,Arsenault_2017,PhysRevE.106.025312}. Currently, the ACTest toolkit has been integrated with the ACFlow toolkit~\cite{Huang:2022}. Thus, it allows direct calls of various methods as implemented in the ACFlow toolkit for analytic continuation calculations.

The rest of this paper is organized as follows. In Section~\ref{sec:overview}, the technical details inside the ACTest toolkit are overviewed. Section~\ref{sec:usage} is the major part of this paper. It elaborates on the basic usages of the ACTest toolkit, including installation, scripts, input files, output files, and control  parameters. In Section~\ref{sec:examples}, an illustrative example is presented. It shows how to generate spectral functions and Green's functions, how to invoke the maximum entropy method in the ACFlow toolkit for massive analytic continuation calculations, and how to visualize the benchmark results, by using the ACTest toolkit. Finally, a brief summary is given in Section~\ref{sec:outlook}.

\section{Overview\label{sec:overview}}
\subsection{Major features}

The current version of the ACTest toolkit is v1.0. Its major features are as follows:
\begin{itemize}
\item ACTest can randomly generate any number of $A(\omega)$, as well as corresponding $G(\tau)$ or $G(i\omega_n)$.

\item ACTest employs one or more parameterized Gaussian, Lorentzian, $\delta$-like, rectangular, and Rise-And-Decay peaks to assemble $A(\omega)$. $A(\omega)$ can be either positive definite for fermionic Green's functions, or non-positive definite for bosonic Green's functions and matrix-valued Green's functions. The frequency grid, i.e., $\omega$, can be either linear or non-linear (such as tangent, Lorentzian, and half-Lorentzian grids).

\item ACTest supports fermionic, bosonic, and symmetric bosonic kernels to generate Green's functions on either imaginary time or Matsubara frequency axes. It supports artificial noise, and the noise level is adjustable.

\item ACTest includes a built-in testing dataset, ACT100, which can serve as a relatively fair standard for examining different analytic continuation methods and codes.

\item ACTest is already interfaced with ACFlow, which is a full-fledged and open-source analytic continuation toolkit~\cite{Huang:2022}. ACTest can access various analytic continuation methods in the ACFlow toolkit, launch them to perform analytic continuation calculations, and provide benchmark reports on their accuracy and efficiency. ACTest also provides a plotting script, that can be used to visualize and compare the true and reconstructed spectral functions.

\item ACTest is an open-source software developed in Julia language. It is quite easy to be extended to implement new features or support the other analytic continuation codes, such as Nevanlinna.jl~\cite{10.21468/SciPostPhysCodeb.19} and SmoQyDEAC.jl~\cite{10.21468/SciPostPhysCodeb.39}. Furthermore, ACTest offers extensive documentation and examples, making it user-friendly.
\end{itemize}

In the following text, we will elaborate on the technical details inside ACTest.

\subsection{Grids for Green's functions\label{subsec:grids}}

As mentioned before, Green's functions in quantum many-body physics are usually defined on imaginary time or Matsubara frequency axes. Imaginary time Green's functions $G(\tau)$ and Matsubara Green's functions $G(i\omega_n)$ are related via Fourier transformation~\cite{many_body_book,many_body_book_2016,RevModPhys.68.13}:
\begin{equation}
G(\tau) = \frac{1}{\beta} \sum_n e^{i\omega_n \tau} G(i\omega_n),
\end{equation}
and
\begin{equation}
G(i\omega_n) = \int^{\beta}_0 d\tau\ e^{-i\omega_n \tau} G(\tau),
\end{equation}
where $\beta$ denotes inverse temperature ($\beta \equiv 1/T$). The grids for Green's functions are linear. Specifically, $\tau_i = i \beta/N_{\tau}$, where $N_{\tau}$ denotes number of time slices and $i \in [0,N_{\tau}]$. $\omega_n = (2n+1)\pi/\beta$ for fermions and $2n\pi/\beta$ for bosons, where $n \in [0, N]$ and $N$ means number of Matsubara frequency points.

\subsection{Meshes for spectral functions\label{subsec:meshes}}

\begin{figure}[ht]
\centering
\includegraphics[width=0.7\textwidth]{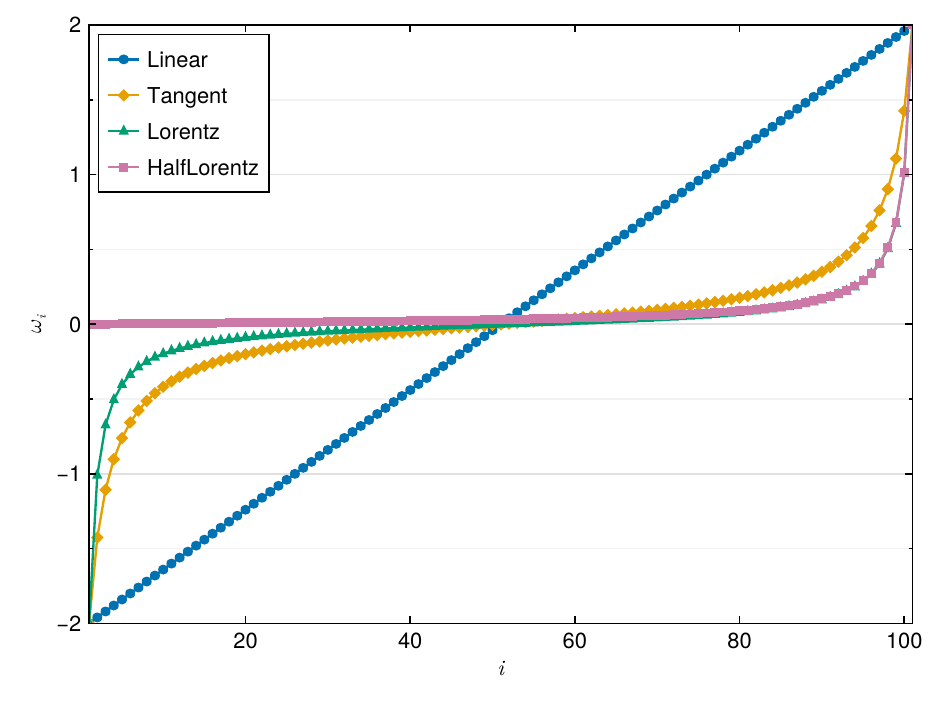}
\caption{The four frequency meshes supported in ACTest: linear, tangent, Lorentzian, and half-Lorentzian meshes. The parameters for these meshes are $\omega_{\text{max}} = -\omega_{\text{min}} = 2.0$ and $N_{\omega} = 101$. \label{fig:mesh}}
\end{figure}

The spectral function $A(\omega)$ is defined on the real frequency axis. The frequency range for $\omega$ is restricted in $[\omega_{\text{min}}, \omega_{\text{max}}]$, where $N_{\omega}$ represents the number of mesh points, $\omega_{\text{min}}$ and $\omega_{\text{max}}$ denote the left and right boundaries of the mesh, respectively. Currently, the ACTest toolkit supports four types of meshes: linear, tangent, Lorentzian, and half-Lorentzian meshes. The latter three meshes are non-linear, with the highest grid density near $\omega = 0$. As its name suggests, the half-Lorentzian mesh is only suitable for cases where $\omega \ge 0$. Figure~\ref{fig:mesh} illustrates typical examples of the four meshes.

\subsection{Peaks\label{subsec:peaks}}

In the ACTest toolkit, the spectral function $A(\omega)$ is treated as a superposition of some peaks (i.e. features). That is to say:
\begin{equation}
A(\omega) = \sum^{N_{p}}_{i = 1} p_i(\omega),
\end{equation}
where $N_{p}$ is the number of peaks, $p(\omega)$ is the peak generation function. Now ACTest supports the following types of peaks:
\begin{itemize}
\item Gaussian peak

\begin{equation}
\label{eq:gauss_peak}
p(\omega) = A\exp{\left[-\frac{(\omega - \epsilon)^2}{2\Gamma^2}\right]}
\end{equation}

\item Lorentzian peak

\begin{equation}
p(\omega) = \frac{A}{\pi} \frac{\Gamma}{(\omega - \epsilon)^2 + \Gamma^2}
\end{equation}

\item $\delta$-like peak

\begin{equation}
p(\omega) = A\exp{\left[-\frac{(\omega - \epsilon)^2}{2\gamma^2}\right]},~
\text{where}~\gamma = 0.01
\end{equation}

\item Rectangular peak

\begin{equation}
p(\omega) =
\begin{cases}
h, \quad \text{if}~\omega \in [c-w/2,c+w/2], \\
0, \quad \text{else}. \\
\end{cases}
\end{equation}

\item Rise-And-Decay peak

\begin{equation}
\label{eq:rise_and_decay_peak}
p(\omega) = h \exp{(-|\omega - c|^{\gamma})}
\end{equation}

\end{itemize}
Here we just use a narrow Gaussian peak to mimic the $\delta$-like peak. In Eq.~(\ref{eq:gauss_peak})-(\ref{eq:rise_and_decay_peak}), $\mathcal{C} = \{A,~\Gamma,~\epsilon,~h,~c,~w,~\gamma\}$ is a collection for essential parameters. The ACTest toolkit will randomize $\mathcal{C}$ and use it to parameterize the peaks.

\subsection{Kernels\label{subsec:kernels}}

Just as stated above, the spectral function and the imaginary time or Matsubara Green's function are related with each other by the Laplace transformation:
\begin{equation}
G(x) = \int d\omega~K(x,\omega) A(\omega).
\end{equation}
Here $K(x,\omega)$ is the so-called kernel function~\cite{PhysRevB.108.235143}. It plays a key role in this equation. In this section, we would like to introduce the kernels that have been implemented in ACTest.

\subsubsection{Fermionic kernels}

For fermionic Green's function, we have
\begin{equation}
\label{eq:ft_kernel}
G(\tau) = \int^{+\infty}_{-\infty} d\omega
          \frac{e^{-\tau\omega}}{1 + e^{-\beta\omega}} A(\omega),
\end{equation}
and
\begin{equation}
G(i\omega_n) = \int^{+\infty}_{-\infty} d\omega
               \frac{1}{i\omega_n - \omega} A(\omega).
\end{equation}
The kernels are defined as
\begin{equation}
K(\tau,\omega) = \frac{e^{-\tau\omega}}{1 + e^{-\beta\omega}},
\end{equation}
and
\begin{equation}
K(\omega_n,\omega) = \frac{1}{i\omega_n - \omega}.
\end{equation}
For fermionic systems, $A(\omega)$ is defined on $(-\infty,\infty)$. It is causal, i.e., $A(\omega) \ge 0$.

\subsubsection{Bosonic kernels}

For bosonic system, the spectral function obeys the following constraint:
\begin{equation}
\text{sign}(\omega) A(\omega) \ge 0.
\end{equation}
It is quite convenient to introduce a new variable $\tilde{A}(\omega)$:
\begin{equation}
\tilde{A}(\omega) = \frac{A(\omega)}{\omega}.
\end{equation}
Clearly, $\tilde{A}(\omega) \ge 0$. It means that $\tilde{A}(\omega)$ is positive definite. So, we have
\begin{equation}
G(\tau)
= \int^{+\infty}_{-\infty} d\omega
          \frac{e^{-\tau\omega}}{1 - e^{-\beta\omega}}
          A(\omega)
=\int^{+\infty}_{-\infty} d\omega
          \frac{\omega e^{-\tau\omega}}{1 - e^{-\beta\omega}}
          \tilde{A}(\omega),
\end{equation}
and
\begin{equation}
G_{B}(i\omega_n) = \int^{+\infty}_{-\infty} d\omega
               \frac{1}{i\omega_n - \omega} A(\omega)
                 = \int^{+\infty}_{-\infty} d\omega
               \frac{\omega}{i\omega_n - \omega} \tilde{A}(\omega).
\end{equation}
The corresponding kernels read:
\begin{equation}
K(\tau,\omega) = \frac{\omega e^{-\tau\omega}}{1 - e^{-\beta\omega}},
\end{equation}
and
\begin{equation}
K(\omega_n,\omega) = \frac{\omega}{i\omega_n - \omega}.
\end{equation}
Especially,
\begin{equation}
K(\tau,\omega = 0) \equiv \frac{1}{\beta},
\end{equation}
and
\begin{equation}
K(\omega_n = 0,\omega = 0) \equiv -1.
\end{equation}

\subsubsection{Symmetric bosonic kernels}

This is a special case for the bosonic Green's function with Hermitian bosonic operators. Here, the spectral function $A(\omega)$ is an odd function. Let us introduce $\tilde{A}(\omega) = A(\omega)/\omega$ again. Since $\tilde{A}(\omega)$ is an even function, we can restrict it in $(0,\infty)$. Now we have
\begin{equation}
G(\tau)
= \int^{\infty}_{0} d\omega
              \frac{\omega [e^{-\tau\omega} + e^{-(\beta - \tau)\omega}]}
                   {1 - e^{-\beta\omega}}
              \tilde{A}(\omega),
\end{equation}
and
\begin{equation}
G(i\omega_n) = \int^{\infty}_{0} d\omega
                   \frac{-2\omega^2}{\omega_n^2 + \omega^2} \tilde{A}(\omega).
\end{equation}
The corresponding kernel functions read:
\begin{equation}
K(\tau,\omega) =
    \frac{\omega [e^{-\tau\omega} + e^{-(\beta - \tau)\omega}]}
    {1 - e^{-\beta\omega}},
\end{equation}
and
\begin{equation}
K(\omega_n, \omega) = \frac{-2\omega^2}{\omega_n^2 + \omega^2}.
\end{equation}
There are two special cases:
\begin{equation}
K(\tau,\omega = 0) = \frac{2}{\beta},
\end{equation}
and
\begin{equation}
\label{eq:bw_kernel_0}
K(\omega_n = 0,\omega = 0) = -2.
\end{equation}

\subsection{Artificial noise\label{subsec:noise}}

Assuming that we already construct the spectral function $ A(\omega) $, it is then straightforward to calculate $ G(\tau) $ or $G(i\omega_n)$ via Eqs.~(\ref{eq:ft_kernel})-(\ref{eq:bw_kernel_0}). At this point, the synthetic Green's function $G$ is exact, containing no numerical noise. We name it $G_{\text{exact}}$. However, the Green's functions obtained from quantum many-body calculations are often noisy~\cite{ASAKAWA2001459,PhysRevB.34.4744,PhysRevLett.55.1204}. This is especially the case in finite-temperature quantum Monte Carlo simulations, where numerical noise is inevitable. To make things worse, when the fermionic sign problem is severe, the noise correspondingly increases. To simulate this scenario, the ACTest toolkit can introduce artificial noise into the synthetic Green's function as follows~\cite{PhysRevB.107.075151}:
\begin{equation}
\label{eq:gnoisy}
G_{\text{noisy}} = G_{\text{exact}}[1 + \delta \mathcal{N}_{C}(0,1)]
\end{equation}
where $\mathcal{N}_{C}(0,1)$ represents complex-valued Gaussian noise with zero mean and unit variance, and the parameter $\delta$ is used to control the noise level ($0 \le \delta \le 1$).

\subsection{Built-in testing dataset}

\begin{figure}[ht]
\centering
\includegraphics[width=0.48\textwidth]{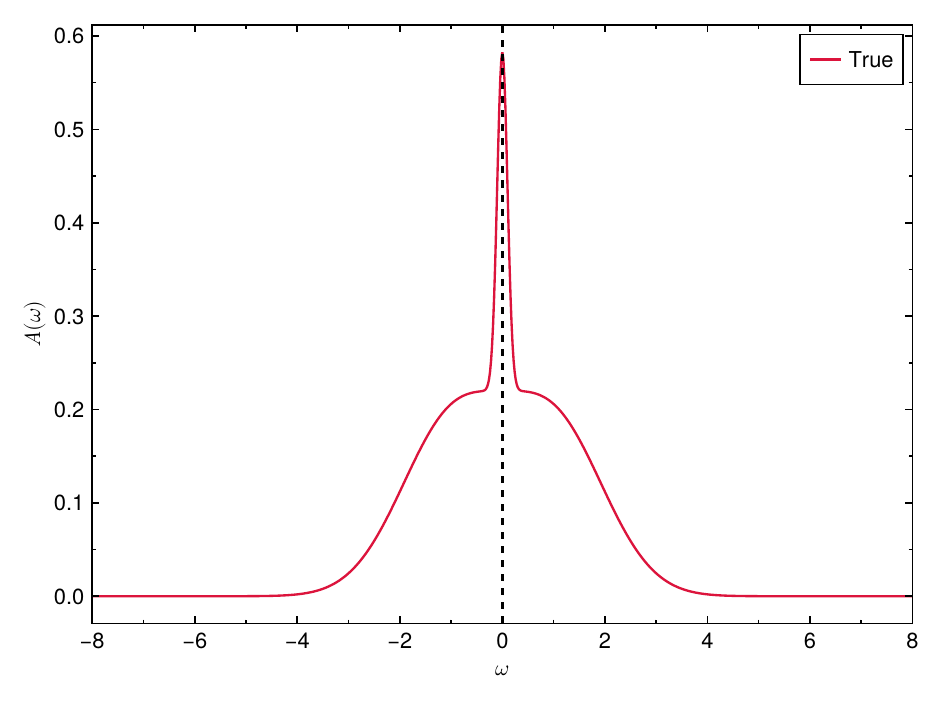}
\includegraphics[width=0.48\textwidth]{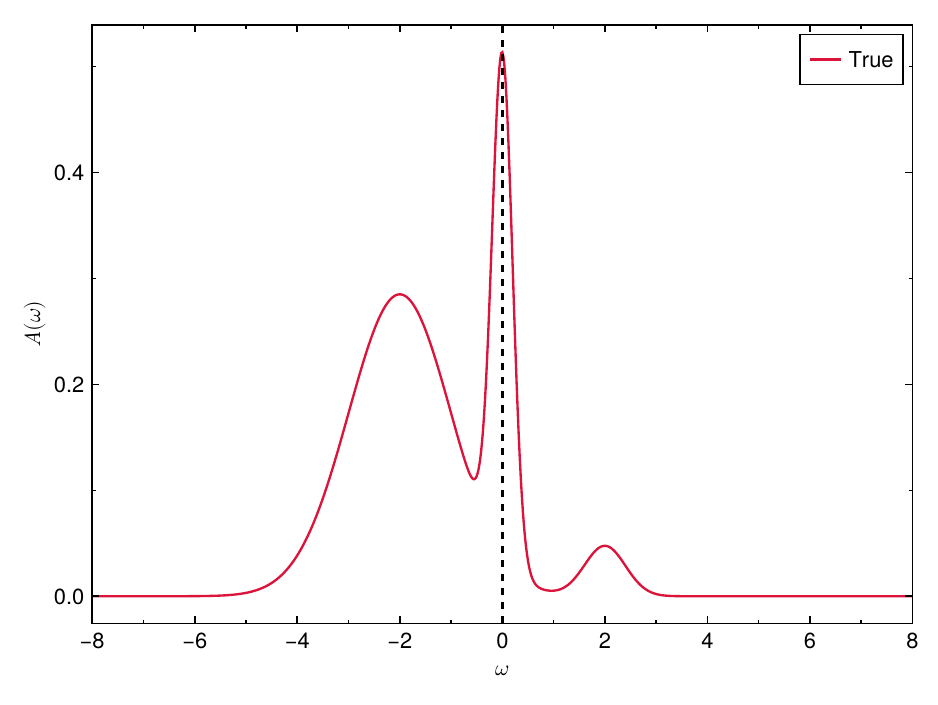} \\
\includegraphics[width=0.48\textwidth]{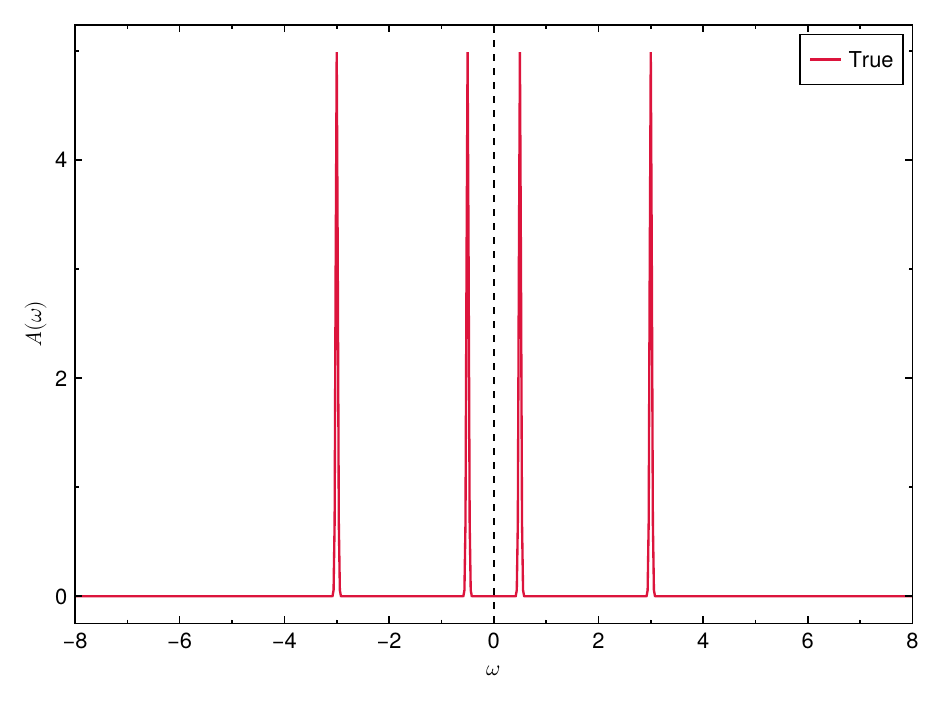}
\includegraphics[width=0.48\textwidth]{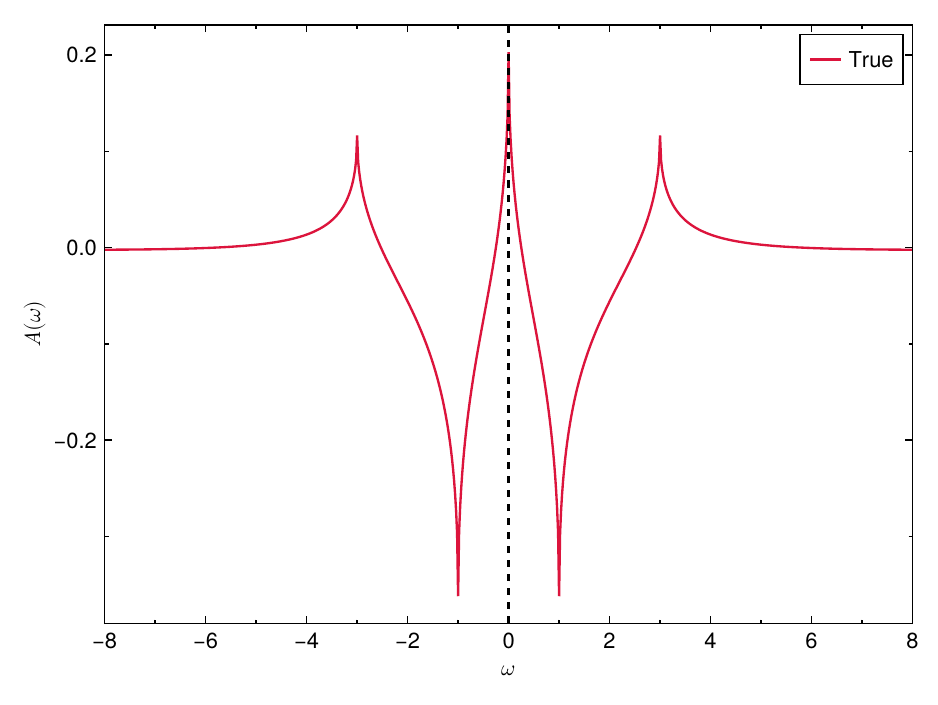}
\caption{Four representative spectra in the ACT100 dataset. The vertical dashed lines denote the Fermi level. The raw data and figures are generated by the \texttt{actest/util/acstd.jl} and \texttt{actest/util/acplot.jl} scripts, respectively. See Section~\ref{subsec:scripts} for more details. \label{fig:act100}}
\end{figure}

The ACTest toolkit operates in two modes. On one hand, ACTest can randomly generate a large number of spectral functions and Green's functions. People can use them to train or examine machine learning models for solving analytic continuation problems. On the other hand, ACTest includes a ``standard'' testing dataset. This dataset comprises 100 representative tests, 50 for fermionic systems and 50 for bosonic systems. Hence we name it the ACT100 dataset. In this dataset, the spectral functions are constructed with predefined parameters (it means that all the peaks and features in these spectra are certain), so they are reproducible. Note that users can only change the grids and the noise levels of the synthetic Green's functions. The basic configurations regarding the fermionic subset of the ACT100 dataset are shown in Table~\ref{tab:act100}. As for the bosonic subset, the basic configurations are identical. The only difference is the use of bosonic kernels to generate the Green's functions. Four typical spectral functions in the ACT100 dataset are illustrated in Figure~\ref{fig:act100}.

\begin{table}[ht]
\centering
\caption{Basic configurations about the fermionic subset in the ACT100 dataset. Clearly, the fermionic subset is classified into three parts. Both the first and the second parts are for diagonal Green's functions, where $A(\omega)$ is positive. Their spectra are continuum and discrete, respectively. The third part is for off-diagonal Green's functions, where $A(\omega)$ is non-positive definite and continuum. The configurations for the bosonic subset are the same. It is also classified into three parts according to positive definiteness of $\tilde{A}(\omega)$ [$\equiv A(\omega)/\omega$]. Full parameters for constructing these spectra are defined in \texttt{actest/src/dataset.jl}. \label{tab:act100}}
\begin{tabular}{l|l|l|l|l}
\hline
\hline
System    & Spectrum's type       & Peak's type    & Number of peaks & Number of spectra / tests \\
\hline
Fermionic & Continuum spectrum,   & Gaussian       & 1               & 3                 \\
          & $A(\omega) \ge 0$     &                & 2               & 7                 \\
          &                       &                & 3               & 10                \\
\hline
Fermionic & Discrete spectrum,    & Gaussian       & 1               & 5                 \\
          & $A(\omega) \ge 0$     &                & 2               & 5                 \\
          &                       &                & 3               & 2                 \\
          &                       &                & 4               & 3                 \\
          &                       &                & 5               & 3                 \\
          &                       &                & 6               & 2                 \\
\hline
Fermionic & Continuum spectrum,   & Rise-And-Decay & 1               & 1                 \\
          & $A(\omega)$ is non-positive definite &                & 2               & 5                 \\
          &                       &                & 3               & 2                 \\
          &                       &                & 4               & 1                 \\
          &                       &                & 5               & 1                 \\
\hline
\hline
\end{tabular}
\end{table}

The ACT100 dataset can be utilized to assess the accuracy of various analytic continuation methods. The question is how to measure the accuracy. Supposed that the true spectral function and the calculated spectral function are $A_{\text{true}}(\omega)$ and $A_{\text{calc}}(\omega)$, respectively, their difference can be calculated by the following expression~\cite{PhysRevB.110.035154}:
\begin{equation}
\label{eq:errA}
\text{Err}(A_{\text{true}},A_{\text{calc}}) = \int^{\omega_{\text{max}}}_{\omega_{\text{min}}} d\omega~\Big{|}A_{\text{true}}(\omega) - A_{\text{calc}}(\omega)\Big{|}~\Big{/}~\int^{\omega_{\text{max}}}_{\omega_{\text{min}}} d\omega~|A_{\text{true}}(\omega)|.
\end{equation}
It is evident that the smaller the value of Err($A_{\text{true}},A_{\text{calc}}$), the better. Thus, the accuracy of an analytic continuation method can be evaluated by:
\begin{equation}
f = \sum^{N_{\text{test}}}_{i = 1}
\theta\left[1 - \text{Err}\left(A^i_{\text{true}},A^i_{\text{calc}}\right)\right]
\left[1 - \text{Err}\left(A^i_{\text{true}},A^i_{\text{calc}}\right)\right].
\end{equation}
Here, $N_{\text{test}}$ is the number of tests, $i$ is the index, $\theta(x)$ is the Heaviside step function. It is obvious that $\text{max}(f) = N_{\text{test}}$ and $\text{min}(f) = 0$.

\subsection{Interface to analytic continuation toolkits}

The ACTest toolkit is not only capable of generating testing datasets, but also of invoking external programs for analytic continuation calculations. Currently, it is integrated with the ACFlow toolkit via the \texttt{actest/util/acflow.jl} script. ACFlow is an open-source toolkit written in Julia~\cite{Huang:2022}. It supports a variety of highly optimized analytic continuation methods, including the maximum entropy method~\cite{JARRELL1996133,PhysRevB.41.2380,PhysRevB.44.6011,PhysRevB.81.155107}, barycentric rational function approximation~\cite{AAA_Lawson,AAA}, Nevanlinna analytical continuation~\cite{PhysRevLett.126.056402,PhysRevB.104.165111}, stochastic analytic continuation (both the Beach's algorithm~\cite{beach2004} and Sanvik's algorithm~\cite{PhysRevX.7.041072,SHAO20231,PhysRevB.57.10287,PhysRevE.94.063308} are supported), stochastic optimization method~\cite{PhysRevB.62.6317,PhysRevB.95.014102}, and stochastic pole expansion~\cite{PhysRevB.108.235143,PhysRevD.109.054508}, etc. The ACTest toolkit is designed in a modular fashion, making it not very difficult to interface with other analytic continuation tools.

\section{Basic usage\label{sec:usage}}

\subsection{Installation}

\subsubsection{Prerequisites}
We chose the Julia language to develop the ACTest toolkit. Julia is an interpreted language. Hence, to run the ACTest toolkit, the latest version of the Julia interpreter must be installed on the target system. Actually, Julia's version number should not be less than 1.6. The core features of the ACTest toolkit rely only on Julia's standard library. However, if one needs to invoke the ACFlow toolkit for analytic continuation calculations, the toolkit must be available in the system. As for how to install and configure the ACFlow toolkit, please refer to the relevant paper~\cite{Huang:2022}. Additionally, if one wants to use the built-in script of ACTest to visualize the calculated results (i.e., the spectral functions), support from the CairoMakie.jl package is necessary. CairoMakie.jl is a 2D plotting program developed in Julia. It can be installed via Julia's package manager:
\begin{verbatim}
    julia> ]
    (@v1.11) pkg> add CairoMakie
\end{verbatim}

\subsubsection{Main program}

The official repository of the ACTest toolkit is hosted on github. Its URL is as follows:
\begin{verbatim}
    https://github.com/huangli712/ACTest
\end{verbatim}
Since the ACTest toolkit has not yet been registered as a regular Julia package, please type the following commands in Julia's REPL to install it:
\begin{verbatim}
    julia> using Pkg
    julia> Pkg.add("https://github.com/huangli712/ACTest")
\end{verbatim}
If this method fails due to an unstable network connection, an offline method should be adopted. First of all, please download the compressed package for the ACTest toolkit from github. Usually it is called \texttt{actest.tar.gz} or \texttt{actest.zip}. Second, please copy it to your favorite directory (such as \texttt{/home/your\_home}), and then enter the following command in the terminal to decompress it:
\begin{verbatim}
    $ tar xvfz actest.tar.gz
\end{verbatim}
or
\begin{verbatim}
    $ unzip actest.zip
\end{verbatim}
Finally, we just assume that the ACTest toolkit is placed in the directory: 
\begin{verbatim}
    /home/your_home/actest
\end{verbatim}
Now we have to manually insert the following codes at the beginning of all ACTest's scripts (see Section~\ref{subsec:scripts}):
\begin{verbatim}
    # Add the ACTest package to the Julia load path
    push!(LOAD_PATH, "/home/your_home/actest/src")
\end{verbatim}
These modifications just ensure that the Julia interpreter can find and import the ACTest toolkit correctly. Now it should work as expected.

\subsubsection{Documentation}

The ACTest toolkit ships with detailed documentation, including the user's manual and application programming interface. They are developed with the Markdown language and the Documenter.jl package. So, please make sure that the latest version of the Documenter.jl package is ready (see \texttt{https://github.com/JuliaDocs/Documenter.jl} for more details). If everything is OK, users can generate the documentation by themselves. Please enter the following commands in the terminal:
\begin{verbatim}
    $ pwd
    /home/your_home/actest
    $ cd docs
    $ julia make.jl
\end{verbatim}
After a few seconds, the documentation is built and saved in the \texttt{actest/docs/build} directory. The entry of the documentation is \texttt{actest/docs/build/index.html}. You can open it with any web browser.

\subsection{Scripts\label{subsec:scripts}}

The ACTest toolkit provides four Julia scripts. They are located in the \texttt{actest/util} directory. Here are brief descriptions for these scripts:

\begin{itemize}
\item \texttt{acgen.jl}

This is the core script of the ACTest toolkit. It is able to randomly generate any number of spectral functions and corresponding Green's functions according to the user's settings.

\item \texttt{acstd.jl}

It is similar to \texttt{acgen.jl}. But its task is to generate the ACT100 dataset, which includes 100 predefined spectral functions and corresponding Green's functions.

\item \texttt{acflow.jl}

This script provides a bridge between the ACTest toolkit and the ACFlow toolkit~\cite{Huang:2022}. At first, it will parse outputs from the \texttt{acgen.jl} or \texttt{acstd.jl} script to get the synthetic Green's functions. Next, these Green's functions are fed into the ACFlow toolkit, which will perform analytic continuation calculations and return the calculated spectral functions. Finally, \texttt{acflow.jl} script will compare the calculated spectral functions with the true solutions, and produce benchmark reports. This script needs support of the ACFlow toolkit. 

\item \texttt{acplot.jl}

It is able to read $A_{\text{calc}}(\omega)$ and $A_{\text{true}}(\omega)$, and plot them in the same figure for comparison. It adopts the PDF format to output the figure. This script requires support of the CairoMakie.jl package.

\end{itemize}
Please execute the above scripts using the following command:
\begin{verbatim}
    $ actest/util/script_name act.toml
\end{verbatim}
Here, \texttt{act.toml} is the configuration file. It is a standard TOML file, primarily used for storing user's settings (i.e., control parameters). The technical details of the \texttt{act.toml} file will be introduced in the following text.

\subsection{Inputs}

The scripts in the ACTest toolkit require only one input file, which is \texttt{act.toml}. Below is a typical \texttt{act.toml} file with some omissions. In this \texttt{act.toml}, text following the ``\#'' symbol is considered as a comment. It contains two sections: \texttt{[Test]} and \texttt{[Solver]}. The \texttt{[Test]} section is mandatory, which controls generation of spectral functions and corresponding Green's functions. We would like to explain the relevant control parameters in the following text. The \texttt{[Solver]} section is optional. Now it is used to configure the analytic continuation methods as implemented in the ACFlow toolkit~\cite{Huang:2022}. Only the \texttt{acflow.jl} script needs to read the \texttt{[Solver]} section. It will transfer these parameters to the ACFlow toolkit to customize the successive analytic continuation calculations. For possible parameters within the \texttt{[Solver]} section, please refer to the documentation of the ACFlow toolkit~\cite{Huang:2022}.
\begin{lstlisting}[language=TOML,
basicstyle=\ttfamily\small,
backgroundcolor=\color{yellow!10},
commentstyle=\color{olive!10!green},
keywordstyle=\color{purple}]
#
# Test: A01
#
# 1. Execute ../../util/acgen.jl act.toml.
# 2. Change solver = "MaxEnt" and modify the [Solver] section accordingly.
# 3. Execute ../../util/acflow.jl act.toml.
# 4. Execute ../../util/acplot.jl act.toml.
#

[Test]
solver  = "MaxEnt"
...

[Solver]
method = "chi2kink"
...
\end{lstlisting}

\subsection{Outputs}

Both \texttt{acgen.jl} and \texttt{acstd.jl} scripts generate \texttt{image.data.i} and \texttt{green.data.i} files. The \texttt{image.data.i} file stores the exact spectral function, i.e., $A_{\text{true}}(\omega)$. The \texttt{green.data.i} file stores the imaginary time Green's function $G(\tau)$ or the Matsubara Green's function $G(i\omega_n)$. The suffix $i$ in filename denotes index for tests. It starts from 1. The \texttt{acflow.jl} script will output quite a few files. The most important one is \texttt{Aout.data.i}. For other possible output files, please refer to the documentation of the ACFlow toolkit~\cite{Huang:2022}. The \texttt{Aout.data.i} file stores the calculated spectral function, i.e., $A_{\text{calc}}(\omega)$. The suffix $i$ in filename also represents index for tests. The aforementioned files are column-based plain texts. They can be opened by any text editor. Their file formats are summarized in Table~\ref{tab:format}. As for the \texttt{acplot.jl} script, it will generate \texttt{image.i.pdf} files, which are image files for $A_{\text{true}}(\omega)$ and $A_{\text{calc}}(\omega)$. Actually, the data from the \texttt{image.data.i} and \texttt{Aout.data.i} files are used to generate the \texttt{image.i.pdf} file.

\begin{table}[ht]
\centering
\caption{File formats for selected files generated by the ACTest toolkit. Here, Var[$G$] means error bar for $G$. Re$G$ and Im$G$ mean real and imaginary parts of $G$, respectively. \label{tab:format}}
\begin{tabular}{l|l|l|l|l|l|l}
\hline
\hline
Filename & Column 1 & Column 2 & Column 3 & Column 4 & Column 5 & Number of lines\\
\hline
\texttt{image.data.i} & $\omega$ & $A_{\text{true}}(\omega)$ & - & - & -  & $N_{\omega}$ \\
\texttt{green.data.i} & $\tau$ & $G(\tau)$ & $\text{Var}[G(\tau)]$ & - & - & $N_{\tau}$ \\
\texttt{green.data.i} & $i\omega_n$ & $\text{Re}G(i\omega_n)$ & $\text{Im}G(i\omega_n)$ & $\text{Var}[\text{Re}G(i\omega_n)]$ & $\text{Var}[\text{Im}G(i\omega_n)]$ & $N$ \\
\texttt{Aout.data.i} & $\omega$ & $A_{\text{calc}}(\omega)$ & - & - & - & $N_{\omega}$ \\
\hline
\hline
\end{tabular}
\end{table}

\subsection{Parameters}

In this subsection, we will provide brief explanations of the control parameters of the ACTest toolkit. The detailed descriptions for these parameters are available in the user's manual. As mentioned before, the \texttt{act.toml} file contains two sections: \texttt{[Test]} and \texttt{[Solver]}. The parameters described here are specific to the \texttt{[Test]} section. They are primarily used for setting generation rules for the spectral functions and Green's functions. However, the parameters within the \texttt{[Solver]} section will be transferred to the ACFlow toolkit for configuring the analytic continuation methods. They will be explained in the documentation of the ACFlow toolkit~\cite{Huang:2022}. 
\begin{itemize}
\item \texttt{solver}

Specify analytic continuation solver. Possible values include ``\texttt{MaxEnt}'', ``\texttt{BarRat}'', ``\texttt{NevanAC}'', ``\texttt{StochAC}'', ``\texttt{StochSK}'', ``\texttt{StochOM}'', and ``\texttt{StochPX}''. They are the analytic continuation methods supported by the ACFlow toolkit. This parameter is only relevant for the \texttt{acflow.jl} and \texttt{acplot.jl} scripts.

\item \texttt{ptype}

Specify type of peaks. Possible values include ``\texttt{gauss}'', ``\texttt{lorentz}'', ``\texttt{delta}'', ``\texttt{rectangle}'', and ``\texttt{risedecay}''. They are corresponding to the Gaussian, Lorentzian, $\delta$-like, rectangular, and Rise-And-Decay peaks, respectively. This parameter is only relevant for the \texttt{acgen.jl} script. See Section~\ref{subsec:peaks} for more details.

\item \texttt{ktype}

Specify kernel function. Possible values include ``\texttt{fermi}'', ``\texttt{boson}'', and ``\texttt{bsymm}''. They are corresponding to the fermionic, bosonic, and symmetric bosonic kernels, respectively. See Section~\ref{subsec:kernels} for more details. 

\item \texttt{grid}

Specify grid for the Green's functions. Possible values include ``\texttt{ftime}'', ``\texttt{btime}'', ``\texttt{ffreq}'', and ``\texttt{bfreq}''. Here, the characters ``\texttt{f}'' and ``\texttt{b}'' mean fermionic and bosonic, respectively. The strings ``\texttt{time}'' and ``\texttt{freq}'' mean imaginary time and Matsubara frequency axes, respectively. See Section~\ref{subsec:grids} for more details. 

\item \texttt{mesh}

Specify mesh for the spectral function. Possible values are ``\texttt{linear}'', ``\texttt{tangent}'', ``\texttt{lorentz}'', and ``\texttt{halflorentz}''. They are corresponding to the linear, tangent, Lorentzian, and half-Lorentzian meshes, respectively. See Section~\ref{subsec:meshes} for more details.

\item \texttt{ngrid}

Number of imaginary time or Matsubara frequency points for the Green's function. It denotes $N_{\tau}$ or $N$. See Section~\ref{subsec:grids} for more details.

\item \texttt{nmesh}

Number of mesh points for the spectral function. It is $N_{\omega}$. See Section~\ref{subsec:meshes} for more details.

\item \texttt{ntest}

How many spectral functions and corresponding Green's functions are generated by the ACTest toolkit? It is the size of the testing dataset for analytic continuation methods or codes. 

\item \texttt{wmax}

Right boundary of the real frequency mesh ($\omega_{\text{max}}$). See Section~\ref{subsec:meshes} for more details.

\item \texttt{wmin}

Left boundary of the real frequency mesh ($\omega_{\text{min}}$). See Section~\ref{subsec:meshes} for more details.

\item \texttt{pmax}

Right boundary of features in the spectrum. It is used to restrict the centers of Gaussian and Lorentzian peaks. $\texttt{wmin} < \texttt{pmin} < \texttt{pmax} < \texttt{wmax}$. See Section~\ref{subsec:peaks} for more details. 

\item \texttt{pmin}

Left boundary of features in the spectrum. It is used to restrict the centers of Gaussian and Lorentzian peaks. $\texttt{wmin} < \texttt{pmin} < \texttt{pmax} < \texttt{wmax}$. See Section~\ref{subsec:peaks} for more details. 

\item \texttt{beta}

Inverse temperature of the system $\beta$ ($\equiv 1/T$). It is used to define the imaginary time or Matsubara frequency grids for Green's functions. See Section~\ref{subsec:grids} for more details.

\item \texttt{noise}

The noise level, i.e., the $\delta$ parameter seen in Eq.~(\ref{eq:gnoisy}). See Section~\ref{subsec:noise} for more details.

\item \texttt{offdiag}

Specify whether the spectral function is positive definite or not. If \texttt{offdiag} is true, it implies that the spectral function is not positive definite and the ACTest toolkit will generate off-diagonal Green's function.  

\item \texttt{lpeak}

It is an integer array that sets the number of peaks (features) that the synthetic spectral function may contain. For example: \texttt{lpeak = [1,2,3,4,5]}, then the ACTest toolkit can generate spectral functions with the number of peaks ranging from 1 to 5. This parameter is only relevant for the \texttt{acgen.jl} script.

\end{itemize}

\section{Examples\label{sec:examples}}

In the \texttt{actest/tests} directory, there are seven typical test cases. Users can modify them to meet their requirements. This section will use an independent example to demonstrate the basic usage of the ACTest toolkit.

\subsection{Generating spectra and correlators}

Now let us test the performance of the maximum entropy method~\cite{JARRELL1996133,PhysRevB.41.2380,PhysRevB.44.6011,PhysRevB.81.155107} in the ACFlow toolkit~\cite{Huang:2022}. We just consider four typical scenarios: (1) Fermionic Green's functions, $A(\omega) > 0$; (2) Fermionic Green's functions, $A(\omega)$ is non-positive definite; (3) Bosonic Green's functions, $A(\omega) > 0$; (4) Bosonic Green's functions, $A(\omega)$ is non-positive definite. For each scenario, we apply the ACTest toolkit to randomly generate 100 spectral functions and corresponding Green's functions. The spectral functions are continuum. They are constructed with Gaussian peaks. The number of possible peaks in each $A(\omega)$ ranges from 1 to 6. The synthetic Green's functions are on the Matsubara frequency axis. The number of Matsubara frequency points is 10. The noise level is $10^{-6}$. The \texttt{act.toml} file for scenario (3) is shown below.
\begin{lstlisting}[language=TOML,
basicstyle=\ttfamily\small,
backgroundcolor=\color{yellow!10},
commentstyle=\color{olive!10!green},
keywordstyle=\color{purple}]
[Test]
solver  = "MaxEnt" # Analytic continuation solver in the ACFlow toolkit
ptype   = "gauss"  # Type of peaks
ktype   = "boson"  # Type of kernels
grid    = "bfreq"  # Type of grids
mesh    = "linear" # Type of meshes
ngrid   = 10       # Number of grid points
nmesh   = 801      # Number of mesh points
ntest   = 100      # Number of tests
wmax    = 8.0      # Right boundary of frequency mesh
wmin    = -8.0     # Left boundary of frequency mesh
pmax    = 4.0      # Right boundary of peaks
pmin    = -4.0     # Left boundary of peaks
beta    = 20.0     # Inverse temperature
noise   = 1.0e-6   # Noise level
offdiag = false    # Whether the spectrum is positive
lpeak   = [1,2,3,4,5,6] # Possible number of peaks
\end{lstlisting}
Once the \texttt{act.toml} file is prepared, the following command should be executed in the terminal:
\begin{verbatim}
    $ actest/util/acgen.jl act.toml
\end{verbatim}
Then, the ACTest toolkit will generate the required data in the present directory. Now there are 100 \texttt{image.data.i} and \texttt{green.data.i} files, where $i$ ranges from 1 to 100. These correspond to $A(\omega)$ and $G(i\omega_n)$ for bosonic systems. We can further verify the data to make sure $A(\omega) > 0$. Finally, we have to copy the \texttt{act.toml} file to another directory, change the \texttt{ktype}, \texttt{grid}, and \texttt{offdiag} parameters in it, and then execute the above command again to generate testing datasets for the other scenarios.

\subsection{Analytic continuation simulations}

\begin{figure}[ht]
\centering
\includegraphics[width=0.48\textwidth]{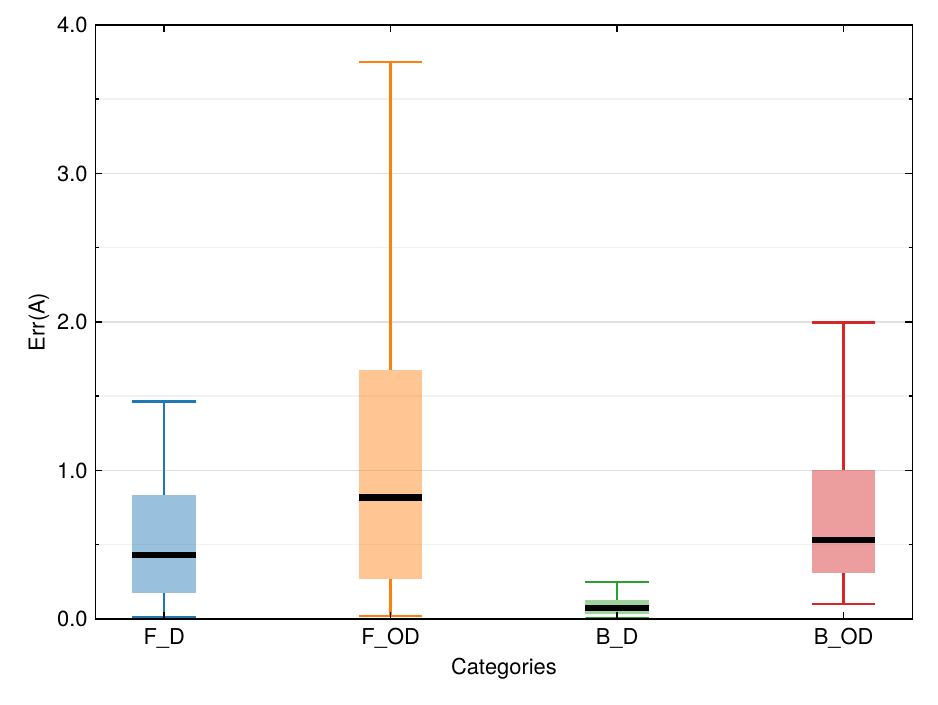}
\includegraphics[width=0.48\textwidth]{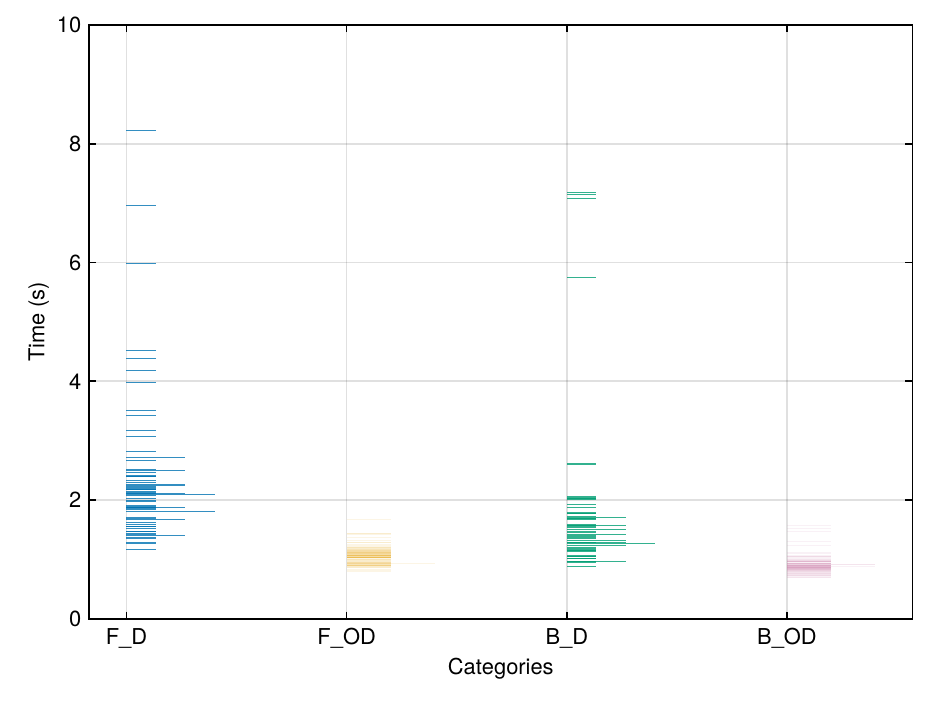}
\caption{Tests of the maximum entropy method as implemented in the ACFlow package~\cite{Huang:2022}. The tests are classified as four categories (scenarios): (1) F\_D, (2) F\_OD, (3) B\_D, and (4) B\_OD. Here, ``F'' means fermionic systems and ``B'' means bosonic systems. ``D'' means diagonal Green's functions [$A(\omega) > 0$] and ``OD'' means off-diagonal Green's functions [$A(\omega)$ is non-positive definite]. Each category contains 100 tests. In this figure, the information contained in the \texttt{summary.data} file is visualized. (Left) Box-and-whisker plot for error statistics. Err(A) is calculated by using Eq.~(\ref{eq:errA}). The black solid segments denote the median values. (Right) Histogram plot for duration times consumed in the tests. \label{fig:performance}}
\end{figure}

Next, we have to add a \texttt{[Solver]} section in the \texttt{act.toml} file to setup the control parameters for the MaxEnt analytic continuation solver (which implements the maximum entropy method) in the ACFlow toolkit~\cite{Huang:2022}. The spirit of the maximum entropy entropy is to figure out the optimal $A$, that minimizes the following functional~\cite{JARRELL1996133}:
\begin{equation}
Q[A] = \frac{1}{2}\chi^2[A] - \alpha S[A].
\end{equation}
Here, $\chi^{2}$ is the so-called goodness-of-fit functional, which measures the distance between the reconstructed Green's function and the original Green's function. $S$ is the entropic term. Usually the Shannon-Jaynes entropy is used. $\alpha$ is a hyperparameter. In this example, we adopt the chi2kink algorithm~\cite{PhysRevE.94.023303} to determine the optimized \(\alpha\) parameters. The list of the $\alpha$ parameters contains 10 elements. The initial value of $\alpha$ is $\alpha_1 = 10^{12}$, and the ratio between two consecutive $\alpha$ parameters ($\alpha_i / \alpha_{i+1}$) is 10. The complete \texttt{[Solver]} section is shown as follows:
\begin{lstlisting}[language=TOML,
basicstyle=\ttfamily\small,
backgroundcolor=\color{yellow!10},
commentstyle=\color{olive!10!green},
keywordstyle=\color{purple}]
[Solver]
method = "chi2kink" # Method to optimize the \alpha parameter
stype  = "sj"       # Typer of the entropic term
nalph  = 12         # Number of \alpha parameter
alpha  = 1e12       # Initial value of \alpha
ratio  = 10.0       # Ratio between two successive \alpha parameters
blur   = -1.0       # Whether we should broaden the kernel
\end{lstlisting}
Please execute the following command in the terminal:
\begin{verbatim}
    $ actest/util/acflow.jl act.toml
\end{verbatim}
It will launch the ACFlow toolkit~\cite{Huang:2022} to perform analytic continuation calculations and generate a lot of output files. In addition to the \texttt{Aout.data.i} file, perhaps the most important file is \texttt{summary.data}. It records the error, status (pass or fail), and duration time for each test. So, we visualize the \texttt{summary.data} file in Figure~\ref{fig:performance}. It is evident that the errors for the analytic continuations of non-diagonal Green's functions [i.e., scenarios (2) and (4)] are slightly larger than those of diagonal Green's functions [i.e., scenarios (1) and (3)]. Furthermore, scenarios (2) and (4) consume much less time to solve the problems. Surprisingly, we found that the pass rates for scenarios (1) and (3) are approximately 80\%, while those for scenarios (2) and (4) are close to 100\%. If we further change the computational configurations (such as increasing the size of the dataset, altering the noise level or changing the type of peaks), and then repeat the aforementioned tests, the final conclusions could be similar.

\subsection{Visualizations}

\begin{figure}[th]
\centering
\includegraphics[width=0.48\textwidth]{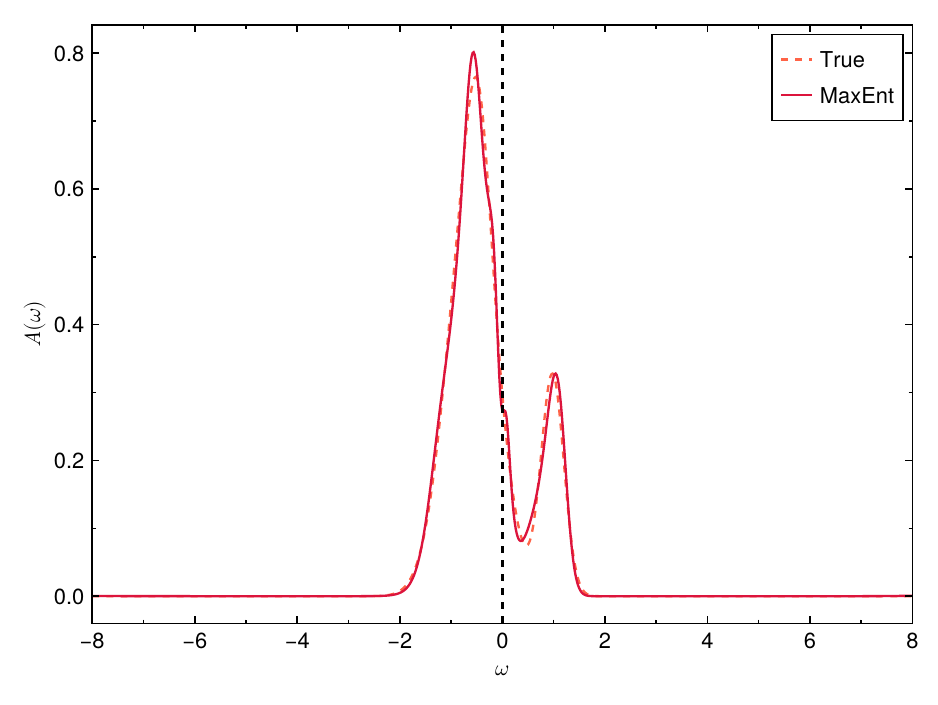}
\includegraphics[width=0.48\textwidth]{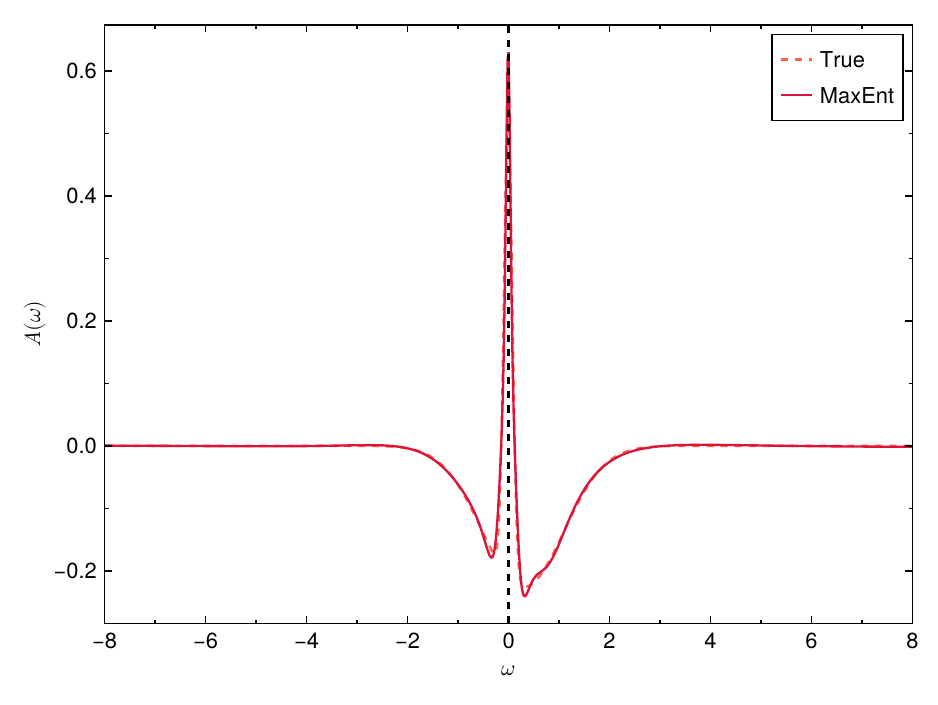}
\caption{Selected analytic continuation results for Matsubara Green's functions (fermionic systems). (Left) Spectra for diagonal Green's function. (Right) Spectra for off-diagonal Green's function. The exact spectra are generated randomly by the ACTest toolkit. The vertical dashed lines denote the Fermi level. \label{fig:fermi}}
\end{figure}

\begin{figure}[th]
\centering
\includegraphics[width=0.48\textwidth]{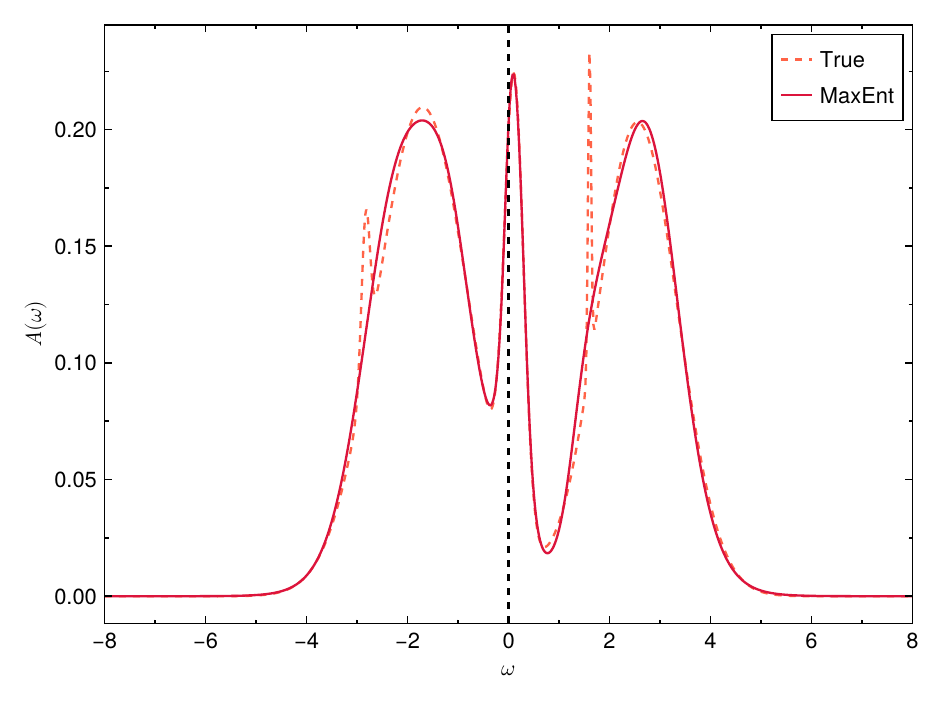}
\includegraphics[width=0.48\textwidth]{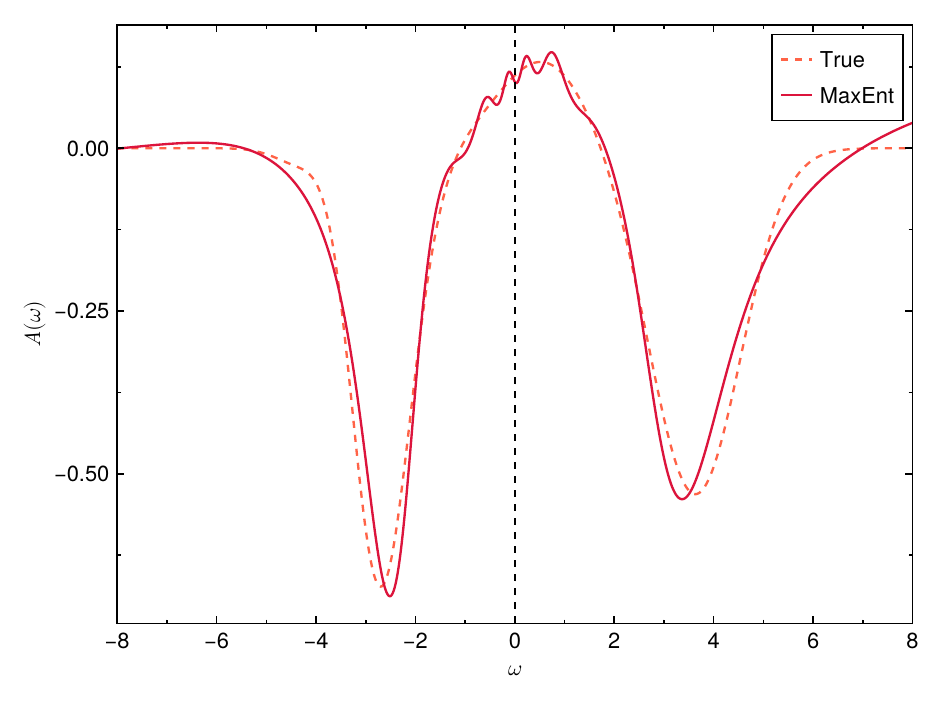}
\caption{Selected analytic continuation results for Matsubara Green's functions (bosonic systems). (Left) Spectra for diagonal Green's function. (Right) Spectra for off-diagonal Green's function. The exact spectra are generated randomly by the ACTest toolkit. The vertical dashed lines denote the Fermi level. \label{fig:boson}}
\end{figure}

Finally, we can utilize the \texttt{acplot.jl} script to generate figures for the spectral functions. To achieve that effect, please execute the following command in the terminal:
\begin{verbatim}
    $ actest/util/acplot.jl act.toml
\end{verbatim}
These figures are in standard PDF format. They are used to analyze the differences between the true spectral functions $A_{\text{true}}(\omega)$ and the calculated spectral functions $A_{\text{calc}}(\omega)$. Figures~\ref{fig:fermi} and ~\ref{fig:boson} illustrate the spectral functions of typical fermionic systems and bosonic systems, respectively. As can be seen from the figures, aside from some very sharp peaks, the calculated spectral functions agree quite well with the true ones.

\section{Concluding remarks\label{sec:outlook}}

In this paper, we introduce the open-source toolkit ACTest, which is designed to generate a large set of spectral functions and corresponding Green's functions for benchmarking analytic continuation algorithms and codes. The toolkit provides a flexible framework to support various grids, meshes, peaks, and kernels. It also supports artificial noise to mimic the realistic Green's functions from quantum many-body computations. The most important feature of ACTest is that it includes a built-in testing dataset, ACT100, which can be used to assess the existing analytic continuation methods in a reproducible and quantitative manner. Now ACTest is interoperable with the ACFlow toolkit, facilitating applications of various analytic continuation methods.

The ACTest toolkit holds promise for significant contributions to the field of solving analytic continuation problems. First of all, it can be used as a validation tool to examine the newly developed analytic continuation methods. Second, if we don't know what kind of analytic continuation problems the present method can deal with, we can turn to ACTest. It will help us to uncover the pros and cons of the present method. Finally, the ACTest toolkit can be used to generate training data for machine learning models. In the future, we would like to extend the ACTest toolkit to support more complex spectral functions and more analytic continuation tools.      

\vspace{10pt}

\noindent \textbf{Declaration of competing interest}

\vspace{5pt}
The author declares that he has no known competing financial interests or personal relationships that could have appeared to influence the work reported in this paper.\\

\noindent \textbf{Data availability}

\vspace{5pt}
The data that support the findings of this study will be made available upon reasonable requests to the corresponding author.\\

\noindent \textbf{Acknowledgement}

\vspace{5pt}
This work is supported by the National Natural Science Foundation of China (under Grants No.~11874329 and No.~11934020).\\

\bibliographystyle{elsarticle-num}
\bibliography{actest}

\end{document}